\documentclass[12pt,arial]{article}

\usepackage[margin=2.0cm]{geometry}
\usepackage[latin1]{inputenc}
\usepackage[spanish,english]{babel}
\usepackage{amsfonts}
\usepackage{amssymb}
\usepackage{graphicx}

\newcommand{\LL}{\mathcal{L}}

\begin{document}

\noindent{\bf PALATINI APPROACH BEYOND EINSTEIN'S GRAVITY}\\
{\bf Gonzalo J. Olmo}, {\small Departamento de F\'{i}sica Te\'{o}rica and IFIC, Centro Mixto Universidad de Valencia \& CSIC. Facultad de F\'{i}sica, Universidad de Valencia, Burjassot-46100, Valencia, Spain.}\\\vspace{14pt}

\noindent{\bf Abstract.}
\noindent I review recent results obtained for extensions of general relativity formulated within the Palatini formalism, an approach in which metric and connection are treated as independent geometrical entities. The peculiar dynamics of these theories, governed by second-order equations and having no new degrees of freedom, makes them specially suitable to address certain aspects of quantum gravity phenomenology, construct nonsingular bouncing cosmologies, and explore black hole interiors, which in the Reissner-Nordstr\"om case develop a compact core of finite density instead of a point-like singularity.\\

\noindent{\bf 1. Introduction}\\ 
\noindent General relativity (GR) has been confronted with experiments in scales that range from millimeters to astronomical distances, scales in which weak and strong field phenomena can be observed \cite{Will-LR}. The theory is so successful in those regimes and scales that it is generally accepted that it should also work at larger and shorter distances, and at weaker and stronger regimes. However, for standard sources of matter and radiation, the theory predicts that the Universe emerged from a singularity and that the fate of sufficiently massive stars is the formation of  black holes, which posses a singularity behind their event horizon. \\
\noindent The general perception is that in such extreme scenarios GR should be replaced by some improved description able to avoid the singularities. This, in particular, has motivated the study of different approaches to the quantization of gravity and also numerous phenomenological extensions of GR. Among the former we find the very famous {\it string theory} \cite{strings} and {\it loop quantum gravity} \cite{LQG}. The phenomenological approaches include theories characterized by higher-order curvature terms and/or higher-order derivatives \cite{f(R)}, models inspired by higher dimensions and the brane-world scenario \cite{branes}, scalar-tensor and scalar-vector-tensor theories \cite{Bruneton:2007si}, and many others. From the existing literature, most of these phenomenological approaches are formulated within the so-called {\it metric approach}, in which the affine connection is defined using the Christoffel symbols of the metric. \\
\noindent A different and, in general, inequivalent approach consists on formulating those theories \`{a} la Palatini, i.e., assuming no a priori relation between the metric and the connection \cite{review}. This possibility is supported by the geometrical nature of gravitation, which follows from the Einstein equivalence principle, and by the fact that metric and connection are independent and fundamental geometrical entities. Therefore, in the construction of extended theories of gravity, Ockham's razor suggests that we should give higher priority to metric-affine theories, in which metric and connection are independent, than to purely metric theories, in which compatibility between metric and connection is implicitly imposed somehow arbitrarily by sociological or educational tradition. In Palatini theories, on the contrary, the connection is determined by solving its corresponding field equation, which is  obtained from the action according to standard variational methods.\\
\noindent In this talk I present recent results obtained within a particular extension of GR formulated \`{a} la Palatini. 
This model allows to explore the potential effects that a minimum length (such as the Planck length) could have on relativistic field theories \cite{Olmo2011}, produces consistent cosmological models that avoid the big bang singularity by means of a cosmic bounce \cite{Bounces}, and modify the internal structure of black holes in such a way that their central singularity is replaced by a compact nucleus that may be nonsingular  \cite{Olmo-Rubiera2011a,Olmo-Rubiera2011b}. \\
 
\noindent {\bf 2. Quantum gravity phenomenology. Introduction of a minimum length.}\\

\noindent The combination of special relativity, quantum theory, and gravity suggests that relativistic quantum gravitational effects could arise at length scales of order $l_P=\sqrt{\hbar G/c^3}\sim 10^{-35}$m. Though this scale is well beyond our current experimental capabilities, its mere existence raises doubts as to how a length, which is not a relativistic invariant, could be consistently introduced in our current field theories to explore the potential phenomenology associated to quantum gravity. \\
\noindent To address this problem, we note that special relativity was built by requiring that the speed of light were an invariant and universal magnitude. To combine the speed of light and the Planck length $l_P$ in a way that preserves the invariant and universal nature of both quantities, we first note that Minkowski space-time allows to interpret the relativity principle in geometrical terms. In this way,  though  $c^2$ has the dimensions of a squared velocity it needs not be seen as a privileged 3-velocity. Rather, it can be regarded as a geometrical invariant in a 4D space-time. Analogously, we may see $l_P^2$ as an invariant with dimensions of length squared in a 4D spacetime.  Dimensional compatibility with a curvature suggests that $l_P^2\equiv 1/R_P$ could be introduced in the theory via the gravitational sector. For this reason, we consider the following action
\begin{equation}\label{eq:f(R)}
S[g_{\mu\nu},\Gamma^\alpha_{\beta\gamma},\psi]=\frac{\hbar}{16\pi l_P^2}\int d^4x \sqrt{-g}\left[R+l_P^2(a R^2+R_{\mu\nu}R^{\mu\nu})\right]+S_m[g_{\mu\nu},\psi] \ ,
\end{equation}
where $R\equiv g^{\mu\nu}R_{\mu\nu}$,  $R_{\mu\nu}\equiv{R^\rho}_{\mu\rho\nu}$ is assumed  symmetric $R_{\mu\nu}=R_{\nu\mu}$ (for the implications of a non-symmetric piece in the Ricci tensor see, for instance, \cite{Ricci_subtlety}), and ${R^\alpha}_{\beta\mu\nu}=\partial_\mu\Gamma_{\nu\beta}^\alpha-\partial_\nu\Gamma_{\mu\beta}^\alpha+\Gamma_{\mu\lambda}^\alpha\Gamma_{\nu\beta}^\lambda-\Gamma_{\nu\lambda}^\alpha\Gamma_{\mu\beta}^\lambda$ represents the components of the Riemann tensor, the field strength of the connection $\Gamma^\alpha_{\mu\beta}$. The field equations for metric and connection that follow from the above action are \cite{review}
\begin{eqnarray}
f_R R_{\mu\nu}-\frac{1}{2}f g_{\mu\nu}+2f_Q R_{\mu\alpha}{R^{\alpha}}_{\nu}&=&\kappa^2 T_{\mu\nu} \label{eq:f(R,Q)-metric}\\
\nabla_\alpha\left[\sqrt{-g}\left(f_R g^{\beta\gamma}+2f_Q R^{\beta\gamma}\right)\right]&=&0 \ . \label{eq:f(R,Q)-connection}
\end{eqnarray}
where $\kappa^2=8\pi l_P^2/\hbar$, $f=R+l_P^2(a R^2+R_{\mu\nu}R^{\mu\nu})$, $f_R\equiv \partial_R f=1+2l_P^2a R$, and $f_Q\equiv \partial_Q f=l_P^2$. Defining the tensor ${P_\mu}^\nu=R_{\mu\alpha}g^{\alpha\nu}$, (\ref{eq:f(R,Q)-metric}) can be seen as a matrix equation, 
\begin{equation}\label{eq:matrix}
2f_Q {P_\mu}^\alpha{P_\alpha}^\nu+f_R{P_\mu}^\nu-\frac{1}{2}f {\delta_\mu}^\nu=\kappa^2 {T_\mu}^\nu \ ,
\end{equation}
which establishes an algebraic relation between the components of ${P_\mu}^\nu$ and those of ${T_\mu}^\nu=T_{\mu\alpha}g^{\alpha\nu}$, i.e., ${P_\mu}^\nu={P_\mu}^\nu({T_\alpha}^\beta)$. 
Once the solution of (\ref{eq:matrix}) is known, the equation for the independent connection can be solved by means of algebraic manipulations. One then finds that this connection can be written as the Levi-Civita connection of a new auxiliary metric ${h}_{\mu\nu}$ (see \cite{OSAT09} for details) which is related to $g_{\mu\nu}$ through the following non-conformal relation
\begin{equation}\label{eq:hmn-general}
{h}^{\mu\nu}=\frac{g^{\mu\alpha}{\Sigma_\alpha}^\nu}{\sqrt{\det \Sigma}} \ ,
\end{equation}
where ${\Sigma_\alpha}^\nu=f_R\delta_\alpha^\nu+2f_Q {P_\alpha}^\nu $ is a function of ${T_\mu}^\nu$ and, therefore, depends on the local densities of energy and momentum.  For instance, if we take the ${T_\mu}^\nu$ of a scalar field with kinetic energy $\chi\equiv g^{\mu\nu}\partial_\mu\phi \partial_\nu\phi$ and Lagrangian $\LL=\chi+2V(\phi)$,  ${h}_{\mu\nu}$ and $g_{\mu\nu}$ turn out to be related by
\begin{equation}
g_{\mu\nu}=\frac{1}{\Omega} {h}_{\mu\nu}\ +\frac{\Lambda_2}{\Lambda_1+\chi\Lambda_2} \partial_\mu\phi \partial_\nu\phi  \label{eq:hdown}
\end{equation}
where $\Omega=\left[\Lambda_1(\Lambda_1+\chi\Lambda_2)\right]^{1/2}$, $\Lambda_1=\sqrt{2f_Q}\lambda+\frac{f_R}{2}$, $\Lambda_2={\sqrt{2f_Q}(-\lambda\pm\sqrt{\lambda^2+\kappa^2 \chi})}/{\chi}$, and $\lambda^2=f/2+f_R^2/8f_Q-\kappa^2\LL/2$. \\
\noindent To better understand the dynamics of our theory, we can use the relation (\ref{eq:hmn-general}) to write the field equation (\ref{eq:f(R,Q)-metric}) in the following compact form
\begin{equation}\label{eq:Rmn-f(R,Q)}
{R_\mu}^\nu({h})=\frac{1}{\sqrt{\det\hat\Sigma}}(\frac{f}{2}{\delta_\mu}^\nu +\kappa^2 {T_\mu}^\nu) \ ,
\end{equation}
where ${R_\mu}^\nu({h})\equiv R_{\mu\alpha}({h}) {h}^{\alpha\nu}$ and ${T_\mu}^\nu\equiv T_{\mu\alpha} g^{\alpha\nu}$. In vacuum (${T_\mu}^\nu=0$) this equation boils down exactly to GR with (possibly) an effective cosmological constant (depending on the form of the Lagrangian). This can be seen by rewriting (\ref{eq:matrix}) in vacuum as 
\begin{equation}
2f_Q\left(\hat{P}+\frac{f_R}{4f_Q}\hat{I}\right)^2=\left(\frac{f^2_R}{8f_Q}+\frac{f}{2}\right)\hat{I}  \ ,
\end{equation}
where $\hat{P}$ and $\hat{I}$ denote the matrices ${P_\mu}^\nu$ and ${\delta_\mu}^\nu$, respectively. The physical solution  to this equation, which recovers the $f(R)$ theory in the limit $f_Q\to 0$, is of the form 
\begin{equation}
{P_\mu}^\nu=-\frac{f_R}{4f_Q}\left(1-\sqrt{1+\frac{4f_Q f}{f_R^2}}\right){\delta_\mu}^\nu\equiv \Lambda(R,Q){\delta_\mu}^\nu  \ .
\end{equation}
This equation can be used to compute $R_0\equiv{P_\mu}^\mu|_{vac}= 4\Lambda(R_0,Q_0)$ and $Q_0={P_\mu}^\alpha {P_\alpha}^\mu|_{vac}=4\Lambda(R_0,Q_0)^2$, which lead to the characteristic relation $Q_0=R^2_0/4$ of de Sitter spacetime. 
For the quadratic models $f(R,Q)=R+aR^2/R_P+Q/R_P$, for instance, one can also use the trace of (\ref{eq:f(R,Q)-metric}) with $g^{\mu\nu}$ to find that $R_0=0$, from which $Q_0=R_0^2/4=0$ follows. For a generic $f(R,Q)$ model, in vacuum  one finds that ${\Sigma_\mu}^\nu= a(R_0){\delta_\mu}^\nu$ and ${h}_{\mu\nu}= a(R_0) g_{\mu\nu}$, with $a(R_0)=f_R\left(1+\sqrt{1+\frac{4f_Q f}{f_R^2}}\right)/2$ evaluated at $R_0$. Therefore, in vacuum (\ref{eq:Rmn-f(R,Q)}) can be written as ${R_\mu}^\nu({h})={R_\mu}^\nu(g)=\Lambda_{eff}{\delta_\mu}^\nu$, with $\Lambda_{eff}=f(R_0,Q_0)/2 a(R_0)^2$, which shows that the field equations coincide with those of GR with an effective cosmological constant.\\

 For the particular model (\ref{eq:f(R)}) with $a=-1/2$ coupled to a scalar field, the low energy-density limit $|\LL/\rho_P|\ll 1$ (where $\rho_P\equiv c^5/8\pi \hbar G^2\sim 10^{94}$g/cm$^3$ is the Planck matter density) leads to
\begin{equation}\label{eq:LowLimit}
R_{\mu\nu}({h})\approx \kappa^2\left(\partial_\mu\phi\partial_\nu\phi+\frac{V}{2}{h}_{\mu\nu}\right) 
+ \frac{1}{\rho_P}\left[({V -{\chi}})\partial_\mu\phi\partial_\nu\phi+\left(\frac{2\kappa^2 V^2+\kappa^2 {\chi}^2}{4}\right){h}_{\mu\nu}\right] 
\end{equation}
which is in agreement with GR up to corrections of order $O(1/\rho_P)$. This indicates that ${h}_{\mu\nu}$ is mainly determined by integrating over the sources (cumulative effects of gravity), whereas $\Omega$ and the last term of (\ref{eq:hdown}) represent local energy-density contributions to the metric.  By neglecting the cumulative effects of gravity, which corresponds to the limit ${h}_{\mu\nu}\approx \eta_{\mu\nu}$, we obtain a kind of  special relativistic limit of the theory (or a DSR-like theory \cite{DSR}). In this limit, the metric becomes
\begin{equation}
g_{\mu\nu}\approx \eta_{\mu\nu}+\frac{2}{\rho_P}\left(V\eta_{\mu\nu}+ \partial_\mu\phi \partial_\nu\phi\right) +O\left(\frac{1}{\rho_P^2}\right). \label{eq:hdown-Mink} 
\end{equation}
From (\ref{eq:hdown-Mink}) we see that the leading order corrections to the Minkowski metric are strongly suppressed by inverse powers of the Planck density, which indicates that a perturbative study of such contributions in field theories is feasible at low energy densities. This should provide an idea of the kind of corrections induced by the Planck-scale modified Palatini dynamics on Minkowskian field theories.  In fact, one can use the metric (\ref{eq:hdown-Mink}) to estimate the first-order modifications of the scalar field equation $\Box \phi -V_\phi=0$ due to the local energy-density dependence of the metric. After some lengthy algebra, one finds
\begin{equation}\label{eq:phi-h}
\partial^2\phi-V_\phi\approx 0+\frac{1}{\rho_P}\left[V_\phi\left(2V -3\partial^\alpha \phi  \partial _{\alpha }\phi\right)+2(\partial^\mu\phi\partial^\nu\phi)\partial_\mu\partial_\nu\phi\right] \ ,
\end{equation}
where $\partial^2\equiv \eta^{\mu\nu}\partial_\mu \partial_\nu$. For a massive scalar with $V(\phi)=m^2\phi^2/2$, the term $V_\phi V$ on the right hand side produces the same effect as a $\lambda \phi^4$ interaction in the Lagrangian with $\lambda\equiv m^4/4\rho_P$. The terms involving derivatives of the field are expected to modify the dispersion relation $E^2=m^2+k^2$ when the scalar amplitude is sufficiently high. This contrasts with other approaches to quantum gravity phenomenology where the proposed modifications of the dispersion relations introduce higher powers of $k^2$ but are independent of the field amplitude. The nonlinear dependence on the field amplitude found here is a distinctive characteristic of Palatini theories, which signals the energy-density dependence of its modified dynamics. \\
     
\noindent {\bf 3. Nonsingular bouncing Palatini cosmologies.}\\      

\begin{figure}[ht]
\begin{center}
{\includegraphics[width=0.8\textwidth]{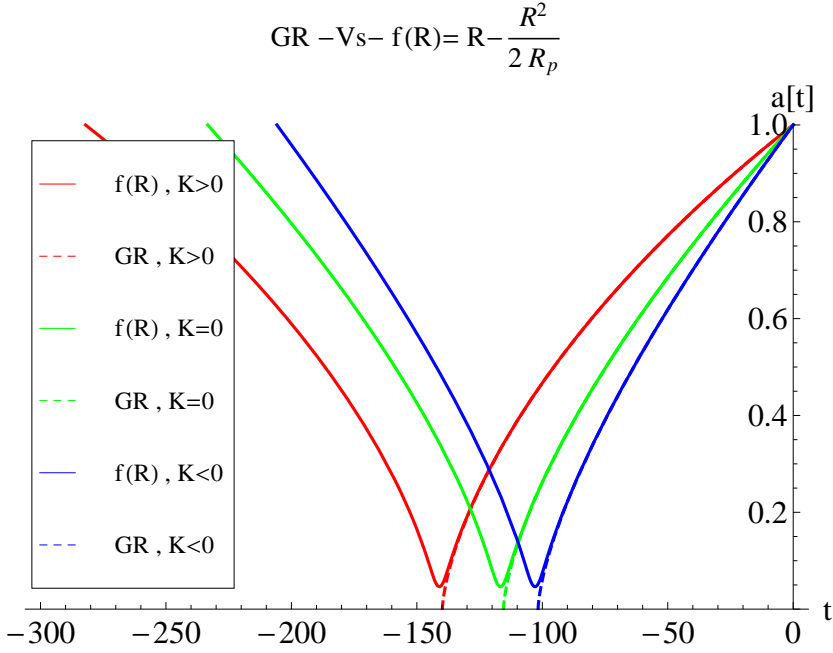}} 
\caption{Time evolution of the expansion factor for the model $f(R)=R-R^2/2R_P$ and $w=0$ for $K>0$, $K=0$, and $K<0$ (solid curves from left to right). From left to right, we see that the universe is initially contracting, reaches a minimum, and then bounces into an expanding phase. The dashed lines, which are only discernible near the bounces, represent the expanding solutions of GR, which begin with a big bang singularity ($a(t)=0$) and quickly tend to the nonsingular solutions.  \label{fig:fR_K_a}}
\end{center}
\end{figure}

\noindent In the previous section we have studied some perturbative properties of Palatini theories. The full dynamics can be explored in simplified scenarios such as cosmological models. In this sense, it is remarkable that 
a simple quadratic Lagrangian of the form $f(R)=R+R^2/R_P$ (where $R_P=1/l_P^2$) does exhibit non-singular solutions \cite{Bounces} for certain equations of state depending on the sign of $R_P$. To be precise, if $R_P > 0$ the bounce occurs for sources with $w=P/\rho> 1/3$. If $R_P < 0$, then the bouncing condition is satisfied by $w < 1/3$ (see Fig.\ref{fig:fR_K_a}). This can be easily understood by having a look at the expression for the Hubble function in a universe filled with radiation plus a fluid with generic equation of state $w$ and density $\rho$
\begin{equation}\label{eq:Hubble-iso}
H^2=\frac{1}{6f_R}\frac{\left[f+(1+3w)\kappa^2\rho+2\kappa^2\rho_{rad}-\frac{6K f_R}{a^2}\right]}{\left[1+\frac{3}{2}\Delta_1\right]^2} 
\end{equation}
where ${\Delta}_1=-(1+w)\rho\partial_\rho f_R/f_R=(1+w)(1-3w)\kappa^2\rho  f_{RR}/(f_R(Rf_{RR}-f_R))$. Due to the structure of $\Delta_1$, one can check that $H^2$ vanishes when $f_R\to 0$. A more careful analysis  shows that $f_R\to 0$ is the only possible way to obtain a bounce with a Palatini $f(R)$ theory that recovers GR at low curvatures if $w$ is constant. In the case of $f(R)=R+R^2/R_P$, it is easy to see that $f_R=0$ has a solution if $1+2R_{Bounce}/R_P=0$ is satisfied for $\rho_{Bounce}>0$, where $R_{Bounce}=(1-3w)\kappa^2\rho_{Bounce}$, which leads to the cases mentioned above. It is worth noting, see Fig.\ref{fig:fR_K_a}, that the expanding branch of the non-singular solution rapidly evolves into the solution corresponding to GR. The departure from the GR solution is only apparent very near the bounce, which is a manifestation of the non-perturbative nature of the solution. Note also that in GR there is a solution that represents a contracting branch that ends at the singularity  where the expanding branch begins (this solution is just the time reversal of the expanding branch). The Palatini model $f(R)=R-R^2/2R_P$ represented here allows for a smooth transition from the initially contracting branch to the expanding one. 
 
\begin{figure}[ht]
\begin{center}
{\includegraphics[width=0.8\textwidth]{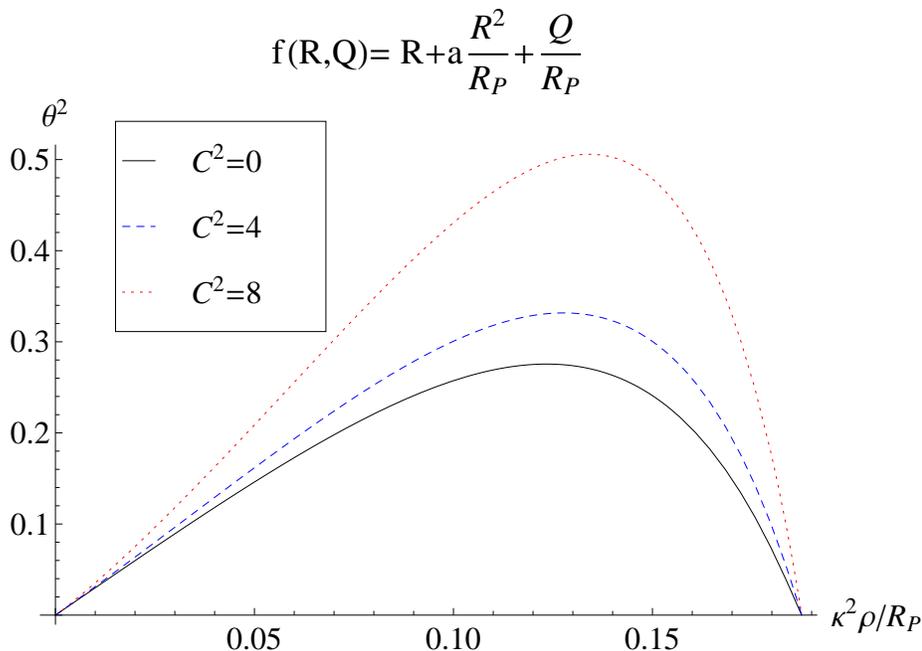}} 
\caption{Evolution of the expansion as a function of $\kappa^2\rho/R_P$ in radiation universes with low anisotropy, which is controlled by the combination $C^2=C_{12}^2+C_{23}^2+C_{31}^2$. The case with $C^2=0$ corresponds to the isotropic flat case, $\theta^2=9H^2$.  \label{fig:ExpanRad}}
\end{center}
\end{figure}

The robustness of the bounce under perturbations  can be tested by studying the solutions of these theories in anisotropic spacetimes of Bianchi-I type
\begin{equation}
ds^2=-dt^2+\sum_{i=1}^3 a_i^2(t)(dx^i)^2 \ .
\end{equation} 
Despite the complexity of this new scenario, one can derive a number of useful analytical expressions for arbitrary Lagrangian of the type $f(R)$. In particular, one finds that the expansion $\theta=\sum_i H_i$ and the shear $\sigma^2=\sum_i\left(H_i-\frac{\theta}{3}\right)^2$ (a measure of the degree of anisotropy) are given by 
\begin{equation}\label{eq:Hubble-f(R)}
\frac{\theta^2}{3}\left(1+\frac{3}{2}\Delta_1\right)^2=\frac{f+\kappa^2(\rho+3P)}{2f_R}+\frac{\sigma^2}{2}
\end{equation}
\begin{equation}\label{eq:shear-f(R)}
\sigma^2=\frac{\rho^{\frac{2}{1+w}}}{f_R^2}\frac{(C_{12}^2+C_{23}^2+C_{31}^2)}{3} \ ,
\end{equation}
where the constants $C_{ij}=-C_{ji}$ set the amount and distribution of anisotropy and satisfy the constraint $C_{12}+C_{23}+C_{31}=0$. In the isotropic case, $C_{ij}=0$, one has $\sigma^2=0$ and $\theta^2=9H^2$, with $H^2$ given by Eq.(\ref{eq:Hubble-iso}). Now, 
since homogeneous and isotropic bouncing universes require the condition $f_R=0$ at the bounce, a glance at (\ref{eq:shear-f(R)}) 
indicates that the shear diverges as $\sim 1/f_R^2$. This shows that, regardless of how small the anisotropies are initially, isotropic $f(R)$ bouncing models with a single fluid characterized by a constant equation of state will develop divergences when anisotropies are present. This negative result, however, does not arise in extended theories of the form (\ref{eq:f(R)}). For that model one finds that $R=\kappa^2(\rho-3P)$, like in GR, and $Q=Q(\rho,P)$ is given by 
\begin{equation}\label{eq:Q}
\frac{Q}{2R_P}=-\left(\kappa^2P+\frac{\tilde f}{2}+\frac{R_P}{8}\tilde f_R^2\right)+\frac{R_P}{32}\left[3\left(\frac{ R}{R_P}+\tilde f_R\right)-\sqrt{\left(\frac{R}{R_P}+\tilde f_R\right)^2-\frac{ 4 \kappa^2(\rho+P)}{R_P} }\right]^2 \ ,
\end{equation}
where $\tilde f=R+aR^2/R_P$, and the minus sign in front of the square root has been chosen to recover the correct limit at low curvatures. In a universe filled with radiation, for which $R=0$, the function $Q$ boils down to   
\begin{equation}
Q= \frac{3R_P^2}{8}\left[1-\frac{8\kappa^2\rho}{3R_P}-\sqrt{1-\frac{16\kappa^2\rho}{3R_P}}\right] \label{eq:Q-rad} \ .
\end{equation}
This expression recovers the GR value at low curvatures, $Q\approx 4(\kappa^2\rho)^2/3+32(\kappa^2\rho)^3/9R_P+\ldots$ but reaches a maximum $Q_{max}=3R_P^2/16$ at $\kappa^2\rho_{max}=3R_P/16$, where the squared root of (\ref{eq:Q-rad}) vanishes. At $\rho_{max}$ the shear also takes its maximum allowed value, namely, $\sigma^2_{max}=\sqrt{3/16}R_P^{3/2}(C_{12}^2+C_{23}^2+C_{31}^2)$, which is always finite, and the expansion vanishes producing a cosmic bounce regardless of the amount of anisotropy (see Fig.\ref{fig:ExpanRad}). Our model, therefore, avoids the well-known problems of anisotropic universes in GR, where anisotropies grow faster than the energy density during the contraction phase leading to a singularity that can only be avoided by sources with $w>1$. 

\noindent {\bf 4. Black holes.} \\

Besides early time cosmology, black hole spacetimes represent another scenario where Palatini theories can be tested. Since the modified dynamics of these theories is induced by the existence of matter sources, vacuum configurations are not suitable for our purposes, because they yield exactly the same solutions as in GR.  In particular, though the Schwarzschild black hole is the most general spherically symmetric, non-rotating vacuum solution of GR and also of (\ref{eq:f(R)}), that solution assumes that all the matter is concentrated on a point of infinite density, which is not consistent with the dynamics of (\ref{eq:f(R)}). In fact, if one considers the collapsing object as described by a perfect fluid that behaves as radiation during the last stages of the collapse, explicit computation of the scalar $Q=R_{\mu\nu}R^{\mu\nu}$ [see (\ref{eq:Q-rad} )] shows that the energy density $\rho$ is bounded from above by $\kappa^2\rho_{max}=3R_P/16$, as we saw in the previous section. Therefore, one should study the complicated dynamical process of collapse of a spherical non-rotating object to determine how the Schwarzschild metric is modified in our theory. For this reason it is easier to study instead vacuum space-times with an electric field, which possess a non-zero stress-energy tensor able to excite the Palatini dynamics even in static settings. The resulting solutions should thus be seen as Planck-scale modifications of the usual Reissner-Nordstr\"om solution of GR.\\

In the context of $f(R)$ theories, electrically charged black holes do not produce any new structures unless one considers that the electromagnetic field is described by some non-linear extension of Maxwell's electrodynamics. This is so because the modified dynamics of $f(R)$ theories is sensitive only to the trace of the energy-momentum tensor of the matter-energy sources, which in the case of Maxwell's theory is zero. For non-linear theories of electrodynamics, like the Born-Infeld model, the corresponding energy-momentum tensor is not traceless and, therefore, is able to produce departures from GR. These black holes have been recently studied in detail in \cite{Olmo-Rubiera2011a}, where it has been shown that exact analytical solutions can be found. In that work one finds that the combination of a quadratic Palatini Lagrangian with Born-Infeld theory can dramatically reduce the intensity of the divergence associated to the singularity. For instance, in GR with Maxwell's electrodynamics one finds that the Kretschmann scalar takes the form  
\begin{equation}\label{eq:Kret-GRM}
R_{\alpha\beta\gamma\delta}R^{\alpha\beta\gamma\delta}=\frac{48 M_0^2}{r^6}-\frac{48 M_0 r_q^2}{r^7}+\frac{14 r_q^4}{r^8} \ ,
\end{equation}
which implies a strong divergence, $\sim 1/r^8$, as $r\to 0$.  In GR coupled to Born-Infeld theory, one finds that the divergence is dominated by  $R_{\alpha\beta\gamma\delta}R^{\alpha\beta\gamma\delta}\sim 1/r^4$. If the gravity Lagrangian is taken as $f(R)=R-l_P^2 R^2$, then the divergence is further reduced to $R_{\alpha\beta\gamma\delta}R^{\alpha\beta\gamma\delta}\sim 1/(r-r_+)^2$, where $r_+>0$ defines the surface of a sphere that contains all the matter and charge of the black hole. Though this solution does not avoid the singularity, it does introduce an important qualitative change with respect to GR, namely, that the matter and charge distribution  of the collapsed object are no longer concentrated on a point, but on a compact sphere. \\

The results of  \cite{Olmo-Rubiera2011a} that we have just summarized suggest that nonperturbative quantum gravitational effects could halt gravitational collapse and produce regular objects sustained by some kind of quantum degeneracy pressure induced by the gravitational interaction, in much the same way as neutron stars and white dwarfs arise when the quantum degeneracy pressure of matter dominates in the interior of stars.  The theory (\ref{eq:f(R)}) does exactly this  \cite{Olmo-Rubiera2011b} . If standard electrically charged black holes are considered under the gravity theory  (\ref{eq:f(R)}), one finds that completely regular solutions exist. These solutions exhibit a compact core of area $A_{core}= N_q\sqrt{2\alpha_{em}} A_P$,  where $A_P=4\pi l_P^2$ is Planck's area, $N_q$ is the number of charges, and  $\alpha_{em}$ is the electromagnetic fine structure constant, which contains all the mass of the collapsed object at a density $\rho_{core}^*=M_0/V_{core}=\rho_P/4\delta_1^*$,  independent of $q$ and $M_0$. The electric charge is distributed on the surface of this core and its density is also a universal constant independent of $q$ and $M_0$, namely, $\rho_q=q/(4\pi r_{core}^2)=(4\pi\sqrt{2})^{-1}\sqrt{c^7/(\hbar G^2)}$.  The impact that these results could have for the theoretical understanding of black holes and the experimental search of compact objects in particle accelerators are currently under investigation. \\

\noindent {\bf 5. Conclusions.} \\
We have shown that a simple extension of general relativity at the Planck scale formulated \`{a} la Palatini successfully addresses different aspects of quantum gravity phenomenology, such as the consistent introduction of a minimum length compatible with the principle of relativity, the avoidance of the big bang singularity, and also the modification of black hole interiors developing a nonsingular compact core that contains all the mass and charge of the collapsed object. In summary, the model (\ref{eq:f(R)}) does everything it was expected to do and lacks of any known instabilities.   \\

\noindent {\bf Acknowledgments.} \\
\noindent Work supported by the Spanish grant FIS2008-06078-C03-02, and the Programme CPAN (CSD2007-00042).

\end{document}